\documentclass[preprint,12pt,natbib209]{emulateapj}

\usepackage{graphicx}
\usepackage{amssymb}
\usepackage{epstopdf}
\usepackage{natbib}
\usepackage{subfigure}
\usepackage{multirow}
\usepackage{verbatim}
\shorttitle{FAR-IR STACKING IN COSMOS}
\shortauthors{LEE ET AL.}

\begin{document}

\title{A Far-Infrared Characterization of 24 $\mu$m Selected Galaxies at $0<z<2.5$ using Stacking at 70 $\mu$m and 160 $\mu$m in the COSMOS Field}
\author{Nicholas Lee\altaffilmark{1}, Emeric Le Floc'h\altaffilmark{1}, D. B. Sanders\altaffilmark{1},D. T. Frayer\altaffilmark{2}, Stephane Arnouts\altaffilmark{3}, Olivier Ilbert\altaffilmark{4}, Herv\'{e} Aussel\altaffilmark{5}, Mara Salvato\altaffilmark{6,7}, N.Z. Scoville\altaffilmark{6}, Jeyhan S. Kartaltepe\altaffilmark{8}}
\altaffiltext{1}{Institute for Astronomy, University of Hawaii, 2680 Woodlawn Dr., Honolulu, HI, 96822, USA}
\altaffiltext{2}{Infrared Processing and Analysis Center, California Institute of Technology 100-22, Pasadena, CA 91125, USA}
\altaffiltext{3}{Canada France Hawaii Telescope Corporation, 65-1238 Mamalahoa Hwy, Kamuela, HI 96743, USA}
\altaffiltext{4}{Laboratoire d'Astrophysique de Marseille, BP 8, Traverse du Siphon, 13376 Marseille Cedex 12, France}
\altaffiltext{5}{UMR AIM (CEA-UP7-CNRS), CEA-Saclay, Orme des Merisiers, b\^{a}t. 709, F-91191 Gif-sur-Yvette Cedex, France}
\altaffiltext{6}{California institute of Technology, MS 105-24, 1200 East California Boulevard, Pasadena, CA 91125, USA}
\altaffiltext{7}{Max Planck Institute for Plasma Physics and Cluster of Excellence, Boltzmann Strasse 2, Garching, D-85748, Germany}
\altaffiltext{8}{National Optical Astronomy Observatory, 950 North Cherry Avenue, Tucson, AZ, 85719, USA}

\begin{abstract}
We present a study of the average properties of luminous infrared galaxies detected directly at 24~$\mu$m in the COSMOS field using a median stacking analysis at 70~$\mu$m and 160~$\mu$m.  Over 35000 sources spanning 0$\leq$$z$$\leq$3 and $0.06$ mJy$\leq$$S_{24}$$\leq$3.0 mJy are stacked, divided into bins of both photometric redshift and 24~$\mu$m flux.  We find no correlation of $S_{70}/S_{24}$ flux density ratio with $S_{24}$, but find that galaxies with higher $S_{24}$ have a lower $S_{160}/S_{24}$ flux density ratio.  These observed ratios suggest that 24~$\mu$m selected galaxies have warmer spectral energy distributions (SEDs) at higher mid-IR fluxes, and therefore have a possible higher fraction of active galactic nuclei.   Comparisons of the average $S_{70}/S_{24}$ and $S_{160}/S_{24}$ colors with various empirical templates and theoretical models show that the galaxies detected at 24~$\mu$m are consistent with ``normal'' star-forming galaxies and warm mid-IR galaxies such as Mrk 231, but inconsistent with heavily obscured galaxies such as Arp 220.  We perform a $\chi^{2}$ analysis to determine best fit galactic model SEDs and total IR luminosities for each of our bins.  We compare our results to previous methods of estimating $L_{\rm{IR}}$ and find that previous methods show considerable agreement over the full redshift range, except for the brightest $S_{24}$ sources, where previous methods overpredict the bolometric IR luminosity at high redshift, most likely due to their warmer dust SED.  We present a table that can be used as a more accurate and robust method for estimating bolometric infrared luminosity from 24 $\mu$m flux densities.
\end{abstract}

\keywords{galaxies: evolution - galaxies: high-redshift - galaxies: statistics - infrared: galaxies}

\section{Introduction}
Although rare in the present-day universe, luminous infrared galaxies (LIRGs) were much more numerous in the past, and they may have played a significant role in the evolution of a large fraction of $L>L^{*}$ galaxies \citep{Sanders:1996p245,Blain:2002p2091,Lagache:2004p2113,LeFloch:2005p2157}.  However, their exact contribution is still poorly understood due to two limitations that have plagued deep surveys performed so far: (1)~the difficulty to identify the most obscured and distant of these objects, as well as measure their redshifts \citep[they are often faint at optical wavelengths because of dust extinction,][]{Houck:2005p2186}, and (2) the difficulty to accurately characterize their nature (bolometric luminosity, mass, physical processes powering their energy output).  Furthermore, because of limited sensitivity of current space- (\emph{Spitzer} MIPS) and ground-based (SCUBA, BOLOCAM, MAMBO, AzTEC,...) observations in the far-IR/submillimeter, only a small number of the most luminous of these sources has been studied in detail.  In addition, at high redshifts there are significant limitations due to confusion, which results from the very large instrument beam characterizing current far-IR/submillimeter observations.  

Many previous studies of LIRGs have been based on data obtained with the \emph{Spitzer Space Telescope}, in particular with the Multiband Imaging Photometer \citep[MIPS,][]{Rieke:2004p2277} at 24~$\mu$m, the detector's most sensitive band.  Using extrapolations based on libraries of galactic infrared (IR) spectral energy distributions (SEDs), the observed 24~$\mu$m flux is converted to a bolometric IR luminosity, $L_{\rm{IR}} \equiv L(8$--$1000 \mu$m), which is then used to calculate properties such as instantaneous star formation rate (SFR).  However, at higher redshifts the 24~$\mu$m band probes shorter rest frame wavelengths, probing rest frame 12 $\mu$m at $z \sim 1$, and rest frame 8 $\mu$m at $z \sim 2$.  The typical peak of the IR SED of star-forming galaxies and galaxies containing active galactic nuclei (AGN) falls around 50--200 $\mu$m; at higher redshifts, the 24~$\mu$m band probes wavelengths farther away from the peak of the IR SED and begins to be heavily affected by broad mid-infrared PAH emission and silicate absorption features.   

Observations at longer wavelengths, such as in the \emph{Spitzer} MIPS  70~$\mu$m and 160~$\mu$m bands (which probe rest frame 24~$\mu$m and 54 $\mu$m at $z \sim 2$, respectively), are needed to more accurately characterize the bolometric luminosity, especially at higher redshifts.  However, the MIPS 70~$\mu$m and 160~$\mu$m bands are significantly less sensitive and have worse angular resolution than the 24~$\mu$m band.  This leads to a drastic decrease in the number of sources directly detected at 70~$\mu$m and 160~$\mu$m, and the galaxies that are detected are biased toward the most luminous sources.  Therefore, we use a stacking analysis \citep[as in][]{Dole:2006p1785,Papovich:2007p38} to study the average 70~$\mu$m and 160~$\mu$m flux densities of galaxies detected at 24~$\mu$m.  In using a stacking analysis we lose the ability to study individual galaxies, but find average properties of galaxies that would otherwise be undetectable.  

In this work we explore the average mid- to far-IR flux densities of galaxies detected at 24~$\mu$m and derive a more accurate method to estimate bolometric IR luminosity.  To accomplish this, we measure stacked 70~$\mu$m and 160~$\mu$m flux densities of galaxies detected at 24~$\mu$m in the Cosmic Evolution Survey (COSMOS) field, binned in both redshift and 24~$\mu$m flux.  We use these stacked fluxes to examine the evolution of mid- to far-IR colors of galaxies as a function of luminosity and redshift.  Our stacked fluxes are fit to libraries of galactic IR SED templates, from which we derive an estimate of the average bolometric IR luminosity.  \citet{Papovich:2007p38} carried out a similar study employing stacking at 70~$\mu$m and 160~$\mu$m, but their analysis was limited by area, with a significantly smaller number of sources.  With an area almost 10 times larger, we obtain more reliable statistics and the ability to bin our sources in narrower bins of redshift and flux.  Our stacked fluxes will eventually be merged with Herschel PACS (100 \& 160~$\mu$m), Herschel SPIRE (200--500~$\mu$m), and SCUBA2 data to get the best sampled SEDs of the high-$z$ literature.  

Throughout this work we denote flux density, $f_{\nu}$ in MIPS 24~$\mu$m, 70~$\mu$m, and 160~$\mu$m bands as $S_{24}$, $S_{70}$, and $S_{160}$, respectively.  When calculating rest-frame quantities, we use a cosmology with $\Omega_{m} = 0.3$, $\Lambda = 0.7$, and $H_{0} = 70$ km s$^{-1}$ Mpc$^{-1}$.

\section{Data}

We use data from the \emph{Cosmic Evolution Survey} (COSMOS) field \citep{Scoville:2007p1511}, a $\sim 2$ deg$^{2}$ field centered at right ascension~$10^{\rm{h}}00^{\rm{m}}28^{\rm{s}}\!.6$, declination~$02^{\circ}12'21''\!\!.0$ (J2000) with extensive multiwavelength imaging and spectroscopic coverage.  In this study, we make use of the COSMOS \emph{Spitzer} \citep[S-COSMOS;][]{Sanders:2007p2324} observations, specifically the data taken by MIPS in the 24~$\mu$m, 70~$\mu$m, and 160~$\mu$m bands.  

\subsection{24~$\mu$m Catalog}
The 24~$\mu$m data reduction and source extraction are detailed in \citet{LeFloch:2009p2898}.  The 24~$\mu$m sources were detected using the automatic procedure of the Sextractor software \citep{Bertin:1996p1695}, and the flux densities were measured with multiple iterations of the point-spread function (PSF) fitting technique of the DAOPHOT package \citep{Stetson:1987p1699}.  Following the convention adopted by the \emph{Spitzer Science Center}, a stellar 10,000 K blackbody spectrum was assumed as the reference SED for the 24~$\mu$m flux density measurements.  The final source list is complete to more than 90\% above a 24~$\mu$m flux of $S_{24} \sim 80 \mu$Jy, and according to simulations, is still reliable down to fluxes as faint as 60~$\mu$Jy, despite a lower completeness of 75\%.   The source list we use for our stacking analysis includes all sources with $S_{24} \geq 60 \mu$Jy.

\subsection{70  $\mu$m and 160~$\mu$m Mosaics}

The MIPS 70~$\mu$m and 160~$\mu$m data were reduced and processed by \citet{Frayer:2009p1271}.  In short, the data were reduced using the Germanium Reprocessing Tools (GeRT, version 20060415) and additional specialized scripts developed for processing survey data from the MIPS-Germanium 70  $\mu$m and 160~$\mu$m detectors.  The final 70  $\mu$m and 160~$\mu$m mosaics have an image pixel scale of 4$\arcsec$ and 8$\arcsec$ and point-source noise ($1 \sigma$) of $1.7$ mJy and 13 mJy, respectively, although there are local background fluctuations across the image depending on the local density of sources.   \citet{Frayer:2009p1271} find 1512 sources at 70~$\mu$m, and 499 sources at 160~$\mu$m ($\geq 5.0 \sigma$), but these detections represent the most luminous sources.  In our stacking analysis, we do not treat these sources differently than 24~$\mu$m sources that were not detected at 70  $\mu$m and 160~$\mu$m (see Section \ref{sec:stacking}).

\subsection{Photometric Redshifts}\label{sec:photo-z}
The extensive multiwavelength coverage of the COSMOS field leads to photometric redshifts (hereafter photo-$z$) with an accuracy better than ever achieved in any other field, as detailed in \citet{Ilbert:2009p1150}.  Optical counterparts of the 24~$\mu$m sources were found from correlating the data with the $K_{s}-$band COSMOS catalog of \citet{McCracken:2009p2464}, as detailed in \citet{LeFloch:2009p2898}.  Photo-$z$ were then calculated using fluxes in 30 bands, covering the far-UV at 1550 \AA \ to the mid-IR at 8.0~$\mu$m.  The uncertainties in the photo-z depend primarily on the redshift and apparent $i^{+}$ magnitude of the source, with errors increasing with fainter and more distant galaxies, but a comparison with faint spectroscopic samples in the COSMOS field revealed a dispersion as low as $\sigma_{\Delta z / (1+z_{s})} = 0.06$ for sources with 23 mag $< i^{+}_{\mathrm{AB}} < 25$ mag at $1.5 \lesssim z \lesssim 3$ \citep{Lilly:2007p1714}.  

Approximately 1000 of our 24 $\mu$m sources are also detected in the X-ray by \emph{XMM-Newton} \citep{Brusa:2010p3149}, and for these sources we use the photo-$z$'s derived from \citet{Salvato:2009p3349}, who have the best photometric redshifts ever produced for AGNs.  In all, we have reliable photo-$z$ for 35,797 sources detected at 24~$\mu$m ($\sim$ 92\% of the 38679 total 24~$\mu$m sources).  We do not use the remaining sources without photo-$z$'s in our study, but these sources are generally at the low $S_{24}$ end of our sample, where we have enough sources to perform a meaningful analysis.

\section{Analysis}\label{sec:stacking}
To study the average 70  $\mu$m and 160~$\mu$m properties of the LIRGs in the COSMOS field, we employ a median stacking analysis to overcome the poor sensitivity of the 70  $\mu$m and 160~$\mu$m MIPS detectors.  Stacking of IR emission has proven valuable for studies such as average 24~$\mu$m fluxes in faint galaxies \citep{Zheng:2006p1772} and contributions to the far IR extragalactic background \citep{Dye:2007p1831}.  A stacking analysis of 70  $\mu$m and 160~$\mu$m fluxes of galaxies selected at 24~$\mu$m has been performed by \citet{Papovich:2007p38}.  However, their analysis used data taken in the Extended \emph{Chandra} Deep Field (ECDF-S), which covers 775 arcmin$^{2}$, and includes only 395 sources.  As a result, they do not have significant detections in some of their lower 24~$\mu$m flux bins.  With the COSMOS data, we have almost 100 times as many sources (stacking efficiency goes as $\sim N^{\frac{1}{2}}$), and will be able to make a much more detailed analysis with narrower bins in both redshift and rest-frame mid-IR luminosity.    

We divide our 24~$\mu$m source list into bins of both redshift and $S_{24}$ before stacking.  Redshift bins ensure that the fluxes we stack were emitted at the same rest-frame wavelength so that we probe the same parts of the SED, and $S_{24}$ bins separate galaxies with different IR luminosities.  Our redshift and $S_{24}$ bins were chosen to maximize the number of sources in each bin while providing the best coverage in redshift and flux; Table \ref{tab:counts} lists the bin limits and the corresponding number of sources in each bin.

\subsection{Stacking Methodology}
We begin by taking a $40 \times 40$ pixel ($2.7' \times 2.7'$ at 70~$\mu$m and $5.3' \times 5.3'$ at 160~$\mu$m) cutout centered around each 24~$\mu$m source in a given bin.  For reference, the FWHM  of the 70~$\mu$m mosaic is $18.6''$ and the FWHM of the 160~$\mu$m mosaic is $39''$.  The size of the cutout does not affect the measured average (stacked) flux as long as the cutout encompasses a large enough area to make a local background estimate.  We center each 70~$\mu$m and 160~$\mu$m subimage on the astrometric coordinates of the 24~$\mu$m source using a bilinear cubic interpolation, and then subtract a local background from each subimage; the local background is calculated from pixels exterior to $\sim 1^{\prime}$ and $\sim 2^{\prime}$ at 70~$\mu$m and 160~$\mu$m, respectively.  Before creating our stacked image, we rotate each subimage by $90^{\circ}$ with respect to the previous subimage to reduce the effects of image artifacts in our analysis.

We then ``stack'' these subimages, aligned at the center (on-source) position and calculate a median flux density at each pixel position.  A median stacking analysis is preferable to mean stacking because the median analysis is more stable and robust to small numbers of bright sources.  The main problem with a mean stacking analysis is that it is very sensitive to bright outliers, which contaminate on-source flux measurements and introduce considerable noise from nearby, bright neighbors.  Most mean stacking studies avoid this problem by removing bright sources from all images before stacking, but this technique solves one problem and creates two more.  Removing bright sources introduces a slight bias against more luminous sources and the resultant stacked flux varies based on the exact flux density cutoff chosen; someone who chooses to remove all sources $\ge 4 \sigma$ will measure a different flux than someone who chooses to remove all sources $\ge 3 \sigma$.  The median stacking analysis avoids these problems, but is more difficult to interpret.  From detailed comparisons of median and mean stacking, \citet{White:2007p1} find that in ``a limit where almost all the values in our sample are small compared with the noise \ldots it is straightforward to interpret our median stack measurements as representative of the mean for the population of sources.''  Our sample fits this description, with only $\sim 4$\% of our sources detected at 70~$\mu$m, and less than 2\% of our sources detected at 160~$\mu$m, so we take the results of our median analysis as representative of the mean flux density.

We calculate the flux of our final stacked image using the DAOPHOT-type photometry IDL procedure, APER, with photometry aperture of $35''$ and sky annulus radii of 39$''$--65$''$ at 70~$\mu$m (at 160~$\mu$m, we use an aperture of $48''$ and sky radii of 64$''$--128$''$).  We use these radii in conjunction with the published MIPS aperture corrections for a 10 K blackbody given by the Spitzer Science Center ($1.48$ at 70~$\mu$m and $1.642$ at 160~$\mu$m) to estimate the total flux density from our stacked source.  

\subsection{Uncertainties in Stacking}

We measure an error in our stacked flux densities from the variance in the local background of the stacked image and the uncertainty in the mean sky brightness.  The absolute calibration of the MIPS detector at long wavelengths is $\sim$10\% \citep{Gordon:2007p3103,Stansberry:2007p3026}, and this dominates the errors in our stacked images in all but the noisiest of bins.  Since a median analysis is a ranking measurement, we find an error in the median by sorting each pixel and then measuring the difference between the middle (median) value and the value that is $N^{1/2}$ ranks away from the middle, where $N^{1/2}$ is the Poissonian noise from a bin with $N$ sources.  For all bins, this represents an almost negligible source of error.    

To test for confusion from nearby bright sources, we searched the 24~$\mu$m catalog for nearby sources that would fall within the apertures used to measure the 70~$\mu$m and 160~$\mu$m flux densities.  We find that only $\sim 3\%$ of our sources have a neighbor within the (larger) aperture used to measure the 160~$\mu$m flux.  However, the location of each of these nearby neighbors relative to the target source will not be uniform, which suggests that the contribution of nearby neighbors detected at 24~$\mu$m to our final stacked flux should be negligible.  The fraction of galaxies with neighbors within the 160~$\mu$m aperture is fairly constant in all bins.

Confusion from faint sources can also add uncertainty to flux measurements; galaxy clustering suggests that the confusion from faint sources will generally be more significant near detected sources than at off-source background positions.  The proper method to account for confusion from faint sources is still currently debated, but the uncertainty is expected to be important mostly for data at very long wavelengths, such as in the submillimeter regime, which is generally confusion-limited (H. Dole 2010, private communication).  The COSMOS MIPS data used in our stacking analysis are not confusion-limited, so we expect a negligible contribution from confusion.

\section{Results}

\subsection{Average 70  $\mu$m and 160~$\mu$m Flux Densities}  
\begin{figure*}[!h]
\centering
\subfigure[Clean Detection in bin with $0.6<z<0.8$ and $0.20$~$<$~$S_{24}$/mJy~$<$~$0.50$]{
	\includegraphics[width=3in]{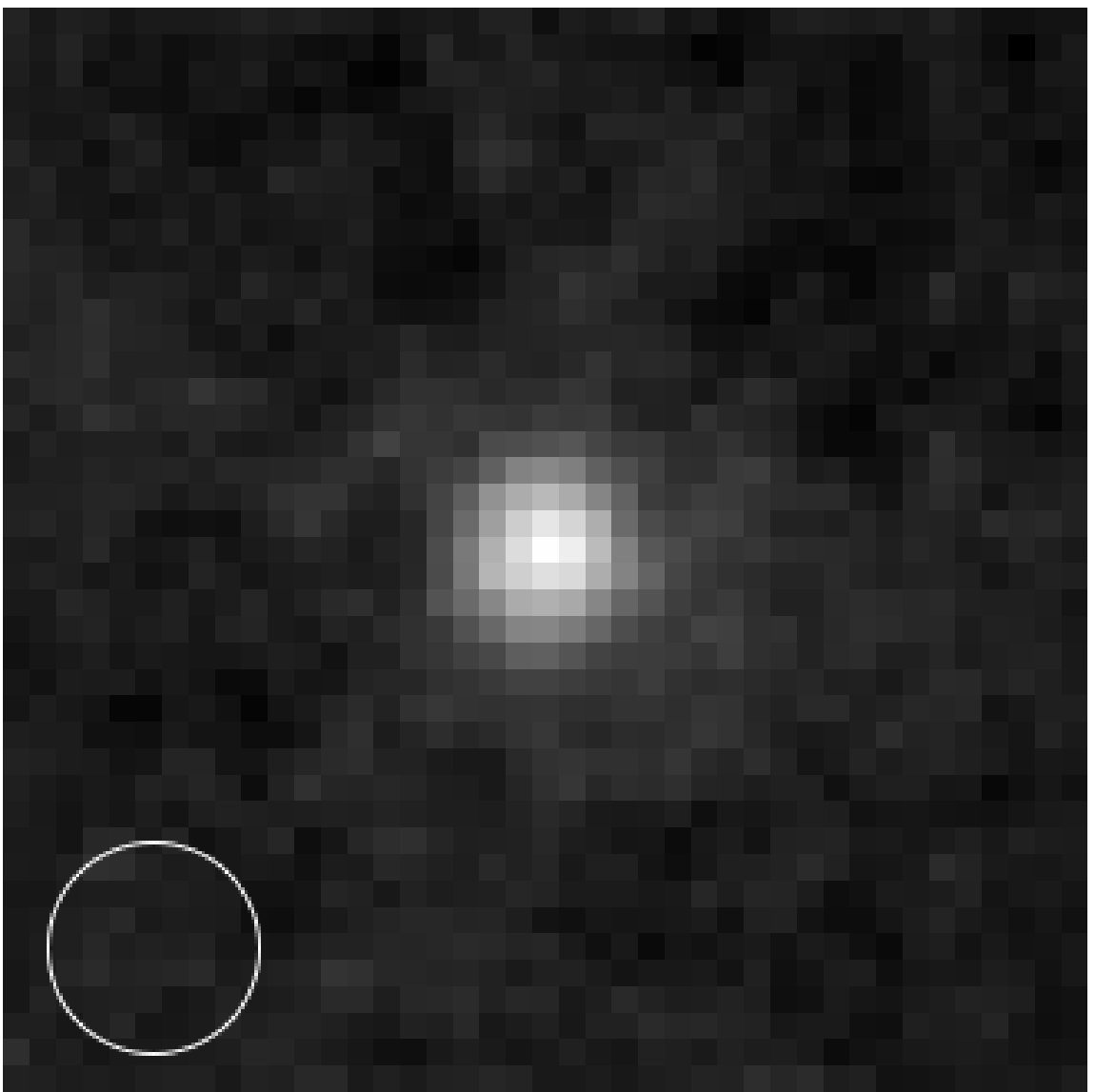}
	\label{fig:goodstack}}
\subfigure[Bad Detection in bin with $1.0<z<1.2$ and $0.06~<~S_{24}/$mJy~$<~0.08$]{
	\includegraphics[width=3in]{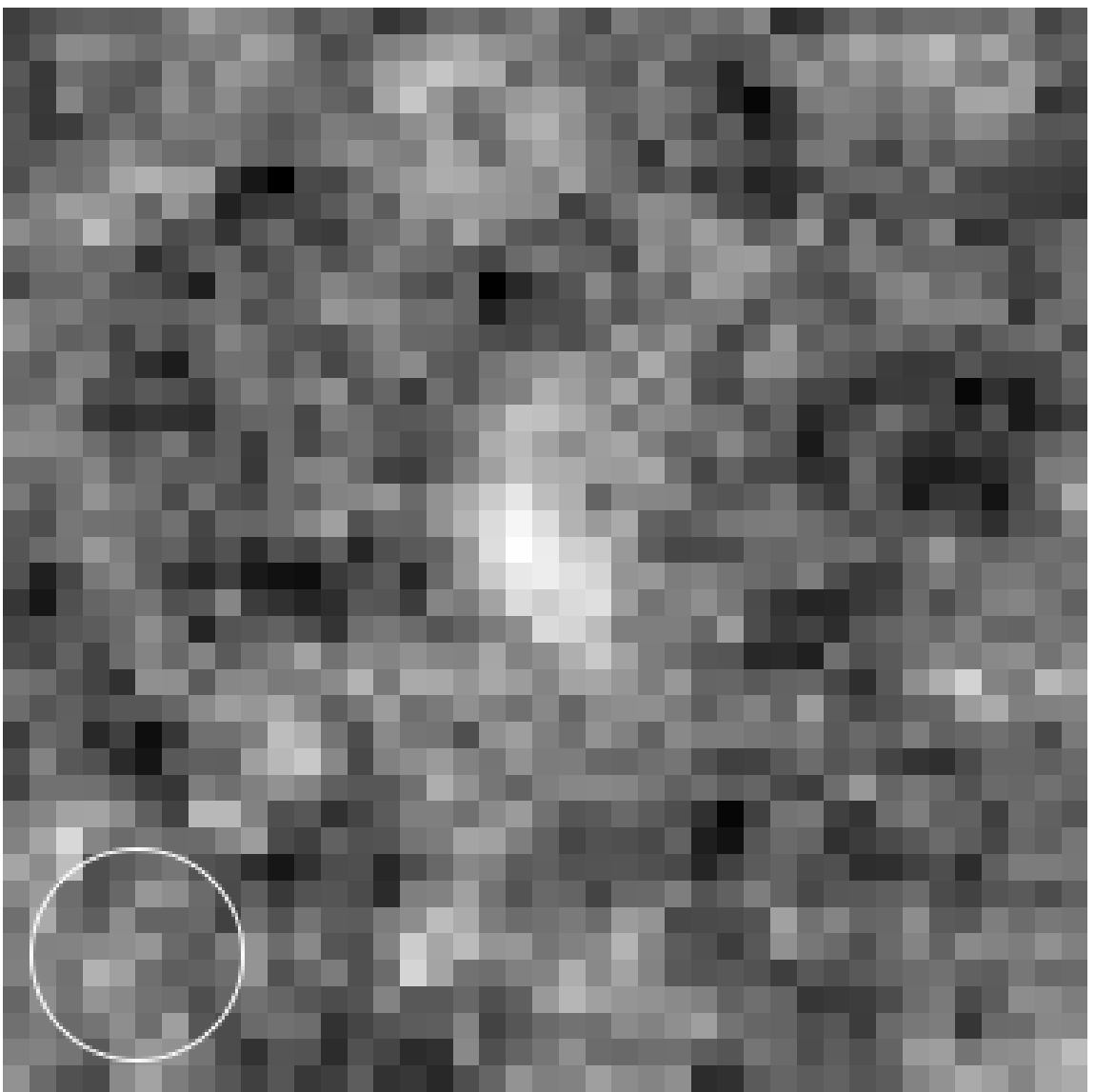}
	\label{fig:badstack}}
\caption{Examples of a clean detection at 70 $\mu$m and a bad detection at 70 $\mu$m.  The bad detection has a source in the center that does not resemble a clean point source, yet aperture photometry of this source yields a signal-to-noise of $\sim 14$.  The circles in the lower left hand corners are the size of the aperture used to measure our stacked fluxes.}
\label{fig:stacks}
\end{figure*}

We performed a median stacking analysis for all 56 bins of redshift and $S_{24}$ in our sample at both 70  $\mu$m and 160~$\mu$m.  From a visual inspection of the images produced by the stacking analysis, we find clean detections in 88\% of our 70~$\mu$m stacks, and 73\% of our 160~$\mu$m stacks.  The rest of the stacks can be split into two categories: (1) non-detections, which have no signal at all, and (2) bad detections, which have a visible, but distorted signal that does not resemble a clean PSF.  The non-detections do not have enough signal to noise for an average source to emerge, but the bad detections do not always have this same problem.  The non-detections and bad detections are mostly in low flux and high redshift bins, although there are a few bad detections in the lowest flux bins with low/intermediate redshifts.  The bins containing the non-detections and bad detections all have a fairly high number of sources, so we believe the lack of a clean detection is due simply to the faintness of the sources we are trying to stack.  In the rest of our analysis, we treat the non-detections as upper limits, but include the bad detections in our full analysis.  Tables \ref{tab:70} and \ref{tab:160} list the measured 70~$\mu$m and 160~$\mu$m fluxes and errors in each of our bins, with upper limits given for non-detections. Figure \ref{fig:stacks} displays an example of a clean detection and a bad detection at 70~$\mu$m.  

\subsection{Evolution of 70/24~$\mu$m and 160/24~$\mu$m Color with 24~$\mu$m Flux}
\label{sec:color}
\begin{figure*}[!h]
\centering
\epsscale{0.8}
\plotone{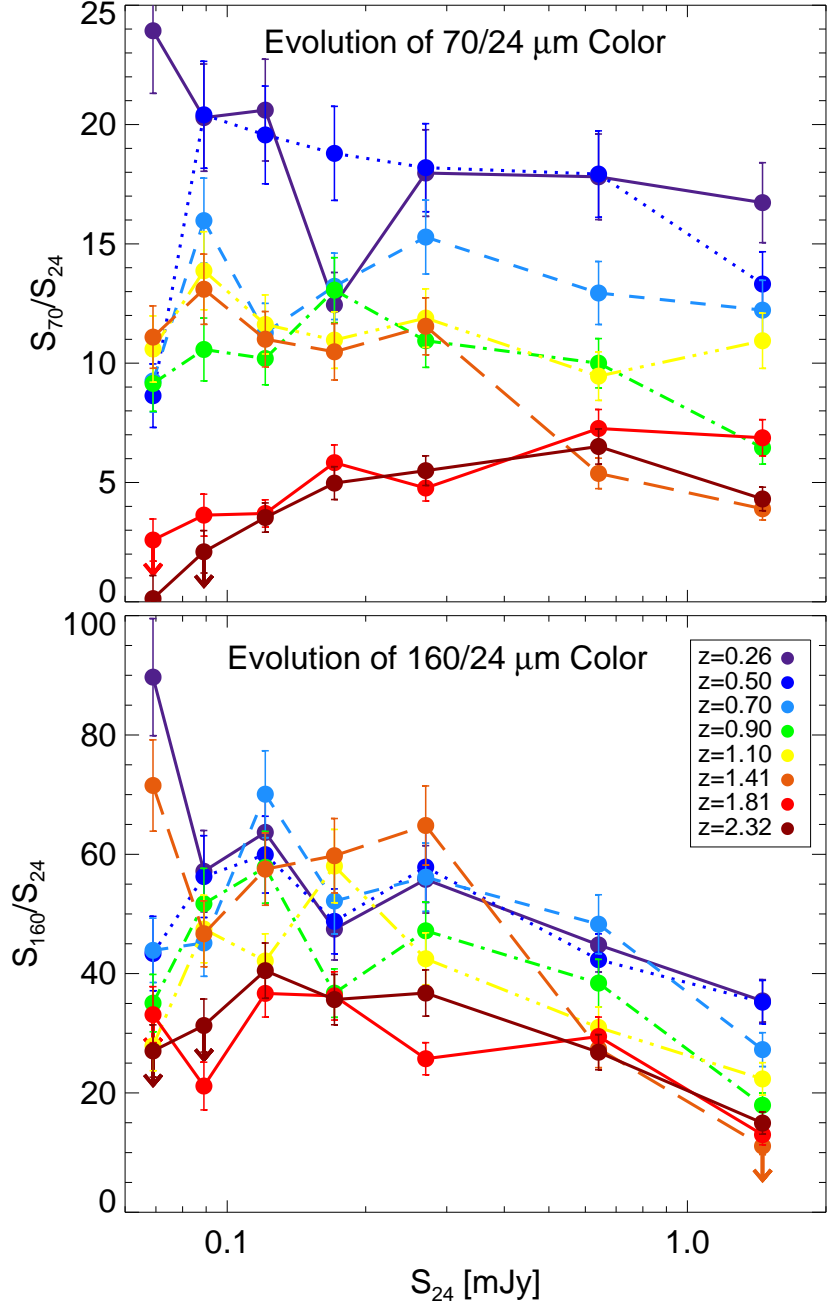}
\caption{Stacked $S_{70}/S_{24}$ (top) and $S_{160}/S_{24}$ (bottom) flux ratios plotted as a function of $S_{24}$.  Each color corresponds to a specific redshift bin, with bluer colors for low redshift and redder colors for high redshift bins.  ``Non-detections'' are marked as upper limits, and 1$\sigma$ error bars are plotted for the rest of the bins.}
\label{fig:compare24}
\end{figure*}

The mid- to far-infrared flux density ratios (or colors) of our stacked galaxies give us insight into the properties of the dust emission from these galaxies.  Figure \ref{fig:compare24} shows the average stacked $S_{70}/S_{24}$ and $S_{160}/S_{24}$ flux ratios of our sources, with each color representing a different redshift bin.  From the top panel of Figure \ref{fig:compare24}, we see that the average $S_{70}/S_{24}$ colors fall in the range $3 \lesssim S_{70}/S_{24} \lesssim 20$, which is roughly consistent with results reported in \citet{Papovich:2007p38}, who found an average $S_{70}/S_{24} \approx 9$.  They also find that sources with $S_{24} > 250 \mu$Jy have a lower average flux ratio ($S_{70}/S_{24} \approx 5$), but we do not see a trend of decreasing $S_{70}/S_{24}$ flux ratio with increasing $S_{24}$.  Our results show a mostly flat $S_{70}/S_{24}$ color with respect to $S_{24}$.  

We also find a trend of decreasing $S_{70}/S_{24}$ flux ratio with increasing redshift (from $S_{70}/S_{24} \approx 15$ at $z \sim 0.3$ to $S_{70}/S_{24} \approx 5$ at $z \sim 2$).  This does not necessarily mean that high redshift galaxies have lower $S_{70}/S_{24}$ flux ratios because at higher redshifts, the observed $S_{70}/S_{24}$ measures flux ratios at shorter rest-frame wavelengths.  Galaxies with strong mid-IR polycyclic aromatic hydrocarbon (PAH) features have mid-IR ($\sim 24 \mu$m) emission that is flatter than their far-IR ($\sim 70 \mu$m) emission, which will lead to a lower observed $S_{70}/S_{24}$ flux ratio at higher redshifts, even when observing galaxies with identical SEDs.  We explore the dependence of $S_{70}/S_{24}$ on redshift further in Section \ref{sec:comparez} through comparisons of our stacking results with models.  

Although we do not see a strong trend in $S_{70}/S_{24}$ ratio with $S_{24}$, we see a clear trend of decreasing $S_{160}/S_{24}$ ratio with increasing $S_{24}$ in the bottom panel of Figure \ref{fig:compare24}.  The average $S_{160}/S_{24}$ ratios we measure range from $10 \lesssim S_{160}/S_{24} \lesssim 100$ and are much larger than the $S_{70}/S_{24}$ ratios because the 160~$\mu$m band samples fluxes emitted at wavelengths closer to the peak of galactic IR SEDs.  \citet{Papovich:2007p38} did not have high enough signal-to-noise in their stacks at 160~$\mu$m to explore $S_{160}/S_{24}$ ratios, but they find average $S_{160}$ ranging from $3.8$ to $10.5$ mJy, broadly consistent with our values of $S_{160}$ in the lower $S_{24}$ bins.      

Figure \ref{fig:compare24} shows that the brightest 24~$\mu$m sources have low $S_{160}/S_{24}$ ratios.  Given that $S_{24}$ broadly correlates with IR luminosity, this means that on average, galaxies with brighter IR luminosities have lower $S_{160}/S_{24}$ flux ratios, and therefore flatter spectra.  This trend is true in all our redshift bins, although the effect is less pronounced in our highest redshift bins.  

The warmer $S_{160}/S_{24}$ colors in the higher $S_{24}$ bins suggest that these sources, on average, have a higher fraction of AGNs \citep{Sanders:1996p245,Laurent:2000p2717}.  Dust grains in the dusty torus around AGN can be heated up to their sublimation temperature (1500~-~2000~K), while dust grains in the diffuse interstellar medium and star-forming regions of galaxies are stochastically heated to lower temperatures around $30\sim40$~K, or up to 200-400~K in HII regions.  The emission from the warmer dust grains around AGN will mostly dominate in the mid-IR wavelengths, while the emission from the colder grains in star-forming regions will dominate the far-IR.  Thus, galaxies powered by AGN will be flatter in their mid-IR to far-IR colors.  

\begin{figure*}[!h]
\centering
\epsscale{0.8}
\plotone{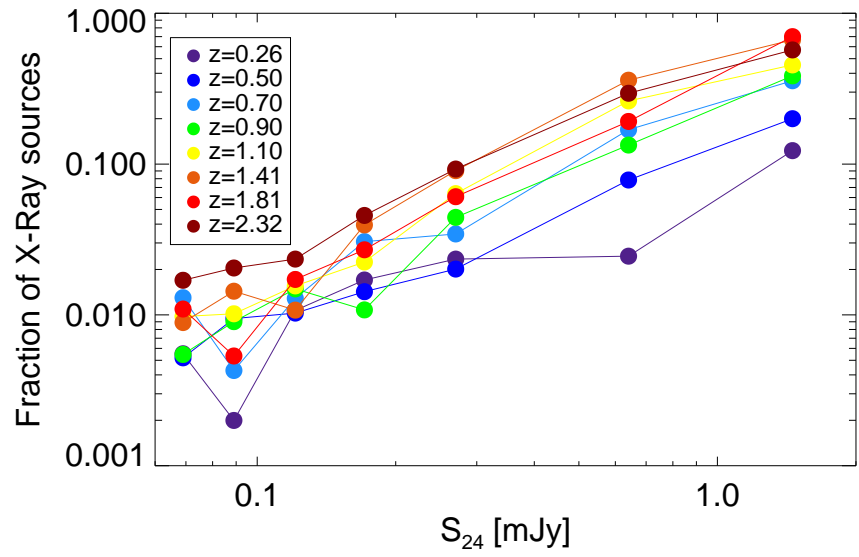}
\caption{Fraction of sources in each bin that are also detected in the X-ray by \emph{XMM-Newton}.  At low $S_{24}$ bins, X-ray sources account for a very small fraction of our sources, but at high flux and high redshift bins the X-ray sources begin to account for a large fraction of sources.}
\label{fig:agn_frac}
\end{figure*}

It should be noted that although we do find a trend of flatter $S_{160}/S_{24}$ ratios that is indicative of a higher fraction of AGN, our sources are still most likely dominated by star formation at the $\sim$85\%--90\% level \citep{LeFloch:2009p2898}.  Out of our over 35000 sources, only $\sim$1000 have X-ray counterparts, suggesting that AGN make up a negligible population of our bins, except for possibly the highest $S_{24}$ bins.  Figure \ref{fig:agn_frac} displays the fraction of sources in each bin that are also detected in the X-ray by \emph{XMM-Newton}.  At dim $S_{24}$ bins, we see that \emph{XMM-Newton} sources indeed account for a low percentage of our sources.  At bright $S_{24}$ bins, the X-ray-detected sources begin to account for an appreciable fraction of our sources, but this does not mean that the mid- and far-IR fluxes of these sources are dominated by AGN.  We discuss this further in Section \ref{sec:SEDfit}.

\section{Discussion}
\subsection{Comparison to Models}\label{sec:comparez}
In this section, we compare the results of our stacking analysis with the expected fluxes and colors from theoretical models and empirical templates.  We first compare our stacked  $S_{70}/S_{24}$ and $S_{160}/S_{24}$ flux ratios with empirical models of ``normal'' star forming galaxies by \citet[][hereafter DALE]{Dale:2002p1836} and models of Arp 220 and Mrk 231 from the SWIRE template library \citep{Polletta:2007p1905}.  We then constrain the average IR luminosity for each bin by fitting to many libraries of theoretical models and empirical templates.

\begin{figure*}[!h]
\centering
\epsscale{0.7}
\plotone{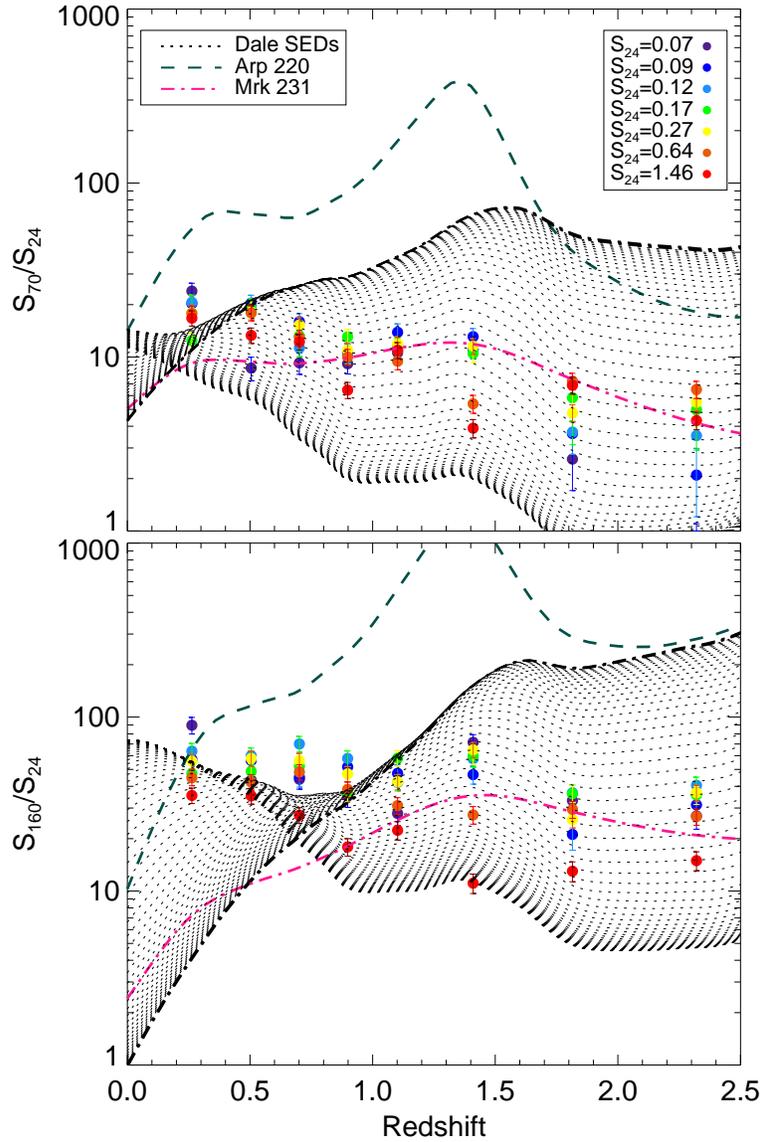}
\caption{Stacked $S_{70}/S_{24}$ and $S_{160}/S_{24}$ flux ratios compared to models of ``normal'' star-forming galaxies from DALE (black lines; note: other empirical models cover a similar range) and models of Arp 220 (dark green line) and Mrk 231 (pink line) from SWIRE.  The DALE models are a one parameter family of models, and we use the full range of DALE models, spanning $1 < \alpha < 2.5$.  The model with the lowest $\alpha$, and also the lowest $L_{\rm{IR}}$, is designated by the dot-dashed line.  Each colored dot represents a different $S_{24}$ bin, with dimmer sources at the blue end of the spectrum and brighter sources represented by redder colors.  We see that our stacked colors are consistent with those of the DALE galaxies and Mrk 231, but not with Arp 220.}
\label{fig:comparez}
\end{figure*}

Figure \ref{fig:comparez} plots $S_{70}/S_{24}$ and $S_{160}/S_{24}$ color as a function of redshift, along with the expected values from the DALE models of star forming galaxies and the SWIRE models of Arp 220 and Mrk 231.  The DALE models are a one parameter family of models and we show models that cover a range of $1 < \alpha < 2.5$, which describe normal star-forming galaxies with $8.3 < log(L_{\rm{IR}}) < 14.3$.  Arp 220 is a well-studied galaxy representative of heavily obscured ULIRGs, while Mrk 231 is representative of galaxies with warm mid-IR colors which are known to host AGN.  We use the code \emph{Le Phare}\footnote{http://www.cfht.hawaii.edu/~arnouts/lephare.html} developed by S. Arnouts and O. Ilbert to determine the flux ratios of the DALE and SWIRE models at varying redshifts.  \emph{Le Phare} is a data analysis package used primarily to compute photometric redshifts, but a preliminary phase of the code also computes theoretical magnitudes, given SED libraries and filter bands.  We can see immediately that the average colors determined from our stacking analysis fall within the region spanned by normal star-forming galaxies and Mrk 231, but our colors do not match those of Arp 220 at any redshift.  This suggests that our 24 $\mu$m selection is biased against heavily obscured objects like Arp 220.  

\subsection{Best-fit Model SEDs and Average Total IR Luminosity}\label{sec:SEDfit}
\begin{figure*}[!h]
\figurenum{5a}
\centering
\includegraphics[width=6in]{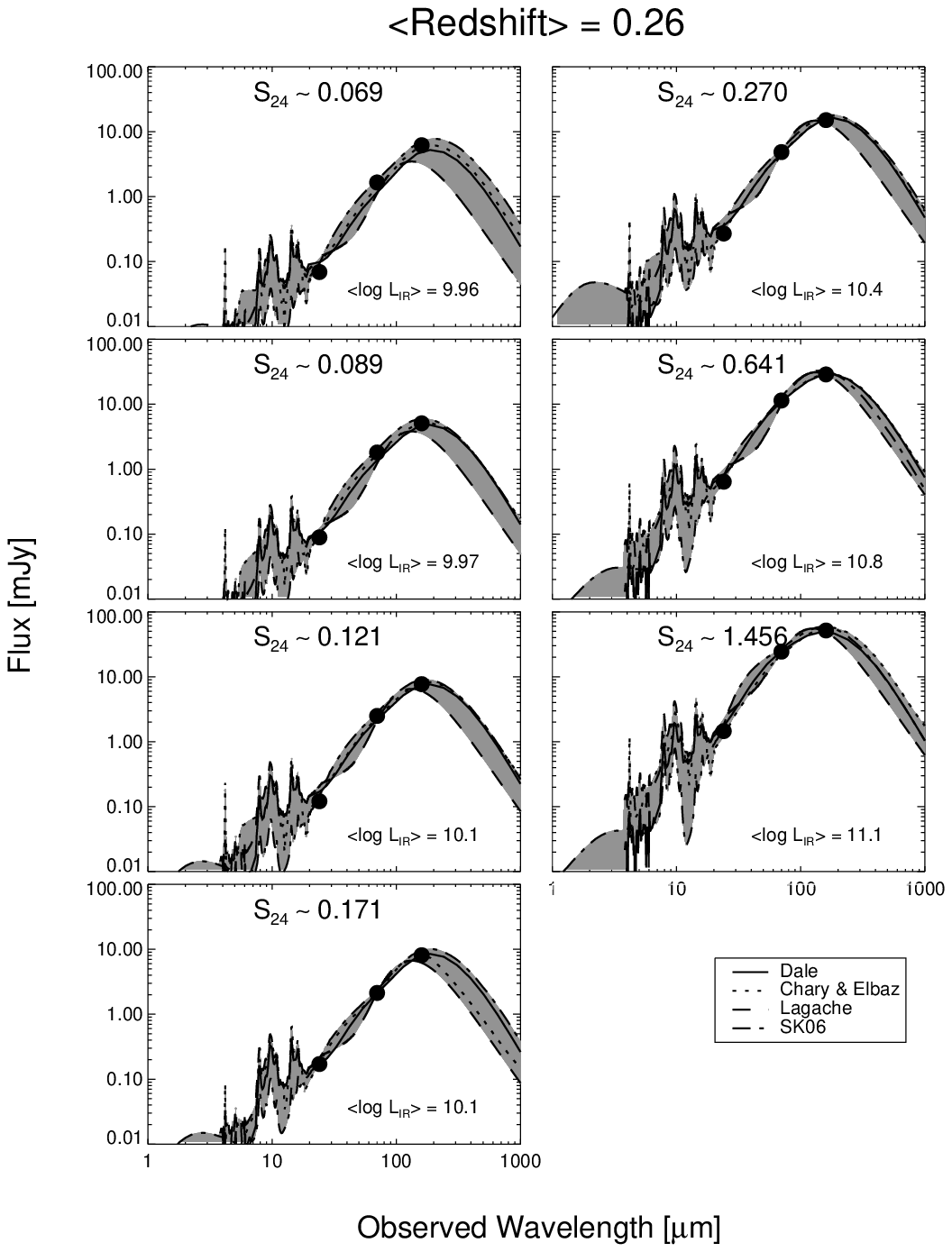}
\caption{SED fits to stacked fluxes in each $S_{24}$ bin at a redshift $0 \leq z \leq 0.4$.  The region spanned by the best-fit SEDs from each library is shaded.  Non-detections are marked as upper limits, and the errors on the rest of the points are smaller than the size of the dot.  Libraries used are from \citeauthor{Dale:2002p1836} (solid line), \citeauthor{Chary:2001p1425} (dotted line), \citeauthor{Lagache:2003p1867} (dashed line), and \citeauthor{Siebenmorgen:2007p1876} (dot-dashed line).  Figures 5b--5h are available in the online version of the Journal.}
\label{fig:matchz0}
\end{figure*}

We use \emph{Le Phare} to perform a $\chi^{2}$ analysis to find best-fit galactic model SEDs for the stacked fluxes calculated in each bin.  We use the empirical templates of \citet{Dale:2002p1836}, \citet{Lagache:2003p1867},  \citet{Chary:2001p1425}, and the theoretical radiation pressure models of \cite{Siebenmorgen:2007p1876}, finding the SED from each library that best fit our stacked data at the correct redshift.   The average $S_{24}$, $S_{70}$, and $S_{160}$ in each bin are plotted along with the best fit models in Figures \ref{fig:matchz0}--5h, arranged by redshift (Figures 5b--5h are only available in the online version of the paper).  Fluxes from the ``non-detcted'' bins are shown as upper limits.  The parameter space spanned by all four of the models is shaded to give an idea of the spread of possible SEDs that fit the data.   

For our model fits, we use four different libraries of ``normal'' star-forming galaxies.  We do have a small fraction of \emph{XMM-Newton}-detected sources that contain AGN, especially in the highest flux bins, but this does not imply that the mid-IR and/or far-IR fluxes of these galaxies is dominated by the AGN itself (many X-ray sources are PAH dominated in the infrared).  Because of the small number of sources, we are unable to perform a separate stacking analysis of only these X-ray sources.  Although we cannot account for the true contribution of AGN contamination, the tight fits we see suggest that most of our sources are indeed star formation dominated.  

From the maximum likelihood function of the $\chi^{2}$ analysis, we estimate a median $L_{\rm{IR}}$ and 1$\sigma$ uncertainties in each bin of our stacking analysis.  We repeat this measurement for each library separately, and then take the mean of the four $L_{\rm{IR}}$ values to estimate the true luminosity.  The dispersion of the luminosities derived from the different libraries is $\sim6\% (3\sigma)$ in all our bins, which suggests that the four libraries are fairly consistent in their estimates of the best-fit $L_{\rm{IR}}$.  We add the 1$\sigma$ uncertainties from each library in quadrature to estimate the error in $L_{\rm{IR}}$, and find typical $3\sigma$ errors around 3\%.  The average IR luminosities  and errors measured for each of our bins are listed in Table \ref{tab:LIR}.  Bins classified as non-detections at both 70~$\mu$m and 160~$\mu$m are marked as upper limits.  As expected, we see that total IR luminosity increases with $S_{24}$ and with redshift.  These galaxies span a large range of IR luminosity, covering ``normal'' galaxies ($L_{\rm{IR}} \leq 10^{11} L_{\sun}$), LIRGs ($10^{11} L_{\sun} \leq L_{\rm{IR}} \leq 10^{12} L_{\sun}$), and ultra luminous infrared galaxies (ULIRGs; $L_{\rm{IR}} \geq 10^{12} L_{\sun}$).  To test our uncertainties, we also found best fit SEDs using 70~$\mu$m and 160~$\mu$m flux densities that were offset by the errors given in Tables \ref{tab:70} and \ref{tab:160} and then measured $L_{\rm{IR}}$ from these fits.  We find discrepancies much less than 6\% from this trial, suggesting that our derived infrared luminosities are robust within the errors in our stacking analysis.  

Table \ref{tab:LIR} gives a relation between \emph{observed} $S_{24}$ and total IR luminosity, with no need for a $k$-correction.  This is an extremely valuable tool given the poor sensitivity of longer wavelength instruments, and provides an effective way to estimate $L_{\rm{IR}}$ when only having 24~$\mu$m data.  Some caution must be taken when using Table \ref{tab:LIR} to estimate $L_{\rm{IR}}$.  Although the agreement between the best-fit models from many libraries is fairly robust, it is not a complete description of the errors.  For most of these bins, we do not have data on the Rayleigh Jean side of our IR SEDs, which means we cannot estimate the cold dust component and its contribution to $L_{\rm{IR}}$.  This means that Table \ref{tab:LIR} is valid under assumption that the libraries used are representative of the diversity of the SEDs beyond $160\mu$m$/(1+z)$.  These results will be tested in the near future with Herschel and SCUBA2.

Figure \ref{fig:matchLIR} shows a comparison of $L_{\rm{IR}}$ calculated from our stacking analysis and $L_{\rm{IR}}$ calculated from extrapolating mid-IR fluxes from only $S_{24}$ based on the Dale SED libraries, as is common practice.  In general, the two methods are in good agreement at low redshifts, but the $L_{\rm{IR}}$ calculated using only 24~$\mu$m data is an overestimate of the true $L_{\rm{IR}}$ at high redshifts, especially for the brighter $S_{24}$ sources.  This is most likely due to the warmer dust SEDs that we find in the bright $S_{24}$ sources (Section \ref{sec:color}).  Our results are consistent with the findings of \citet{Calzetti:2007p1434}, who find that rest-frame 24 $\mu$m flux is a much better indicator of bolometric infrared luminosity than 8 $\mu$m flux, and with the findings of \citet{Papovich:2007p38}, who find that the $L_{\rm{IR}}$ estimated without taking into account stacked 70~$\mu$m and 160~$\mu$m fluxes overestimates the true $L_{\rm{IR}}$.  To summarize, extrapolating a bolometric infrared luminosity from a 24 $\mu$m flux density without taking into account 70~$\mu$m and 160~$\mu$m flux will result in an overestimate of $L_{\rm{IR}}$ at high redshifts.  Table \ref{tab:LIR} will give a more accurate and robust estimate of $L_{\rm{IR}}$ at these redshifts.  

\section{Summary}
We perform a median stacking analysis on over 35000 sources detected directly at 24~$\mu$m in the COSMOS field at 0$\leq$$z$$\leq$3 and $0.06$ mJy$\leq$$S_{24}$$\leq$3.0 mJy to study their average flux densities at 70~$\mu$m and 160~$\mu$m.  Of the 56 bins used, 95\% had detections at 70~$\mu$m and 93\% had detections at 160~$\mu$m.  Analysis of the $S_{70}/S_{24}$ and $S_{160}/S_{24}$ flux density ratios suggest the following.
\begin{itemize}
\item 24~$\mu$m sources have average flux-density ratios consistent with empirical models of ``normal'' star-forming galaxies or with warm mid IR galaxies, like Mrk 231, which are known to host AGN. 
\item 24~$\mu$m sources have average flux-density ratios that are inconsistent with Arp 220, which suggests that 24~$\mu$m is not very useful for finding heavily obscured objects like Arp 220.
\item Sources with brighter $S_{24}$ have warmer $S_{160}/S_{24}$ flux ratios, decreasing by a factor of 2 from $0.1 \lesssim S_{24}/\rm{mJy} \lesssim 1.0$, which implies that galaxies with brighter infrared luminosities have a higher fraction of AGN.
\end{itemize}
Our stacking analysis provides the largest statistical study of the average far-IR flux densities of the faint 24~$\mu$m population.  A comparison of the average far-IR fluxes to libraries of empirical templates and theoretical models allows us to estimate the total IR luminosity of a typical galaxy detected at 24 $\mu$m within certain redshift and $S_{24}$ bins.  We find that previous studies based on extrapolating $L_{\rm{IR}}$ from 24 $\mu$m data, without far-IR stacking, generally overpredict the total infrared luminosity, especially at higher redshifts.  A more accurate method for estimating $L_{\rm{IR}}$ using only 24 $\mu$m flux and redshift is provided in Table \ref{tab:LIR}, which takes into account the average mid- and far-IR fluxes of 24~$\mu$m selected galaxies.  

\acknowledgments It is a pleasure to acknowledge the contribution from all our colleagues of the COSMOS collaboration.  More information on the COSMOS survey is available at http://www.astro.caltech.edu/cosmos.  This work is based on observations made with the \emph{Spitzer Space Telescope}, a facility operated by NASA/JPL.  Financial supports were provided by NASA through contracts 1289085, 1310136, 1282612, and 1298231 issued by the Jet Propulsion Laboratory.  We are grateful to Herve Dole for insightful discussions on stacking techniques.

\begin{deluxetable*}{ccccccccc}
\tablecaption{Number of Sources in Each Redshift and $S_{24}$ Bin \label{tab:counts}}
\tablehead{\colhead{$S_{24}$} & \multicolumn{8}{c}{Redshift Range} \\
\colhead{(mJy)} & \colhead{0--0.4} & \colhead{0.4--0.6} & \colhead{0.6--0.8} & \colhead{0.8--1.0} & \colhead{1.0--1.2} & \colhead{1.2--1.6} & \colhead{1.6--2} & \colhead{2--3} } 
\startdata
0.06--0.08 & 724 & 578 & 923 & 1281 & 1025 & 1573 & 1006 & 1002 \\
0.08--0.10 & 501 & 423 & 701 & 887 & 688 & 1045 & 749 & 684 \\
0.10--0.15 & 841 & 681 & 1011 & 1479 & 1092 & 1395 & 1280 & 896 \\
0.15--0.20 & 528 & 421 & 588 & 834 & 535 & 557 & 629 & 482 \\
0.20--0.50 & 1152 & 745 & 930 & 1354 & 628 & 597 & 790 & 604 \\
0.50--1.00 & 408 & 153 & 136 & 187 & 61 & 75 & 73 & 78 \\
1.00--3.00 & 228 & 35 & 28 & 26 & 11 & 21 & 20 & 21 \\
\enddata
\end{deluxetable*}

\begin{deluxetable*}{ccccccccc}
\tabletypesize{\footnotesize}
\tablecaption{Average 70~$\mu$m Flux Densities [mJy] and Errors \label{tab:70}}
\tablehead{\colhead{$S_{24}$} & \multicolumn{8}{c}{Redshift Range} \\
\colhead{(mJy)} & \colhead{0--0.4} & \colhead{0.4--0.6} & \colhead{0.6--0.8} & \colhead{0.8--1.0} & \colhead{1.0--1.2} & \colhead{1.2--1.6} & \colhead{1.6--2} & \colhead{2--3} } 
\startdata
0.06--0.08 & 1.65{\scriptsize{$\pm$0.18}} & 0.60{\scriptsize{$\pm$0.09}} & 0.64{\scriptsize{$\pm$0.09}} & 0.63{\scriptsize{$\pm$0.08}} & 0.73{\scriptsize{$\pm$0.10}} & 0.77{\scriptsize{$\pm$0.09}} & $<$0.18{\scriptsize{$\pm$0.06}} & $<$0.01{\scriptsize{$\pm$0.07}} \\
0.08--0.10 & 1.81{\scriptsize{$\pm$0.20}} & 1.82{\scriptsize{$\pm$0.20}} & 1.42{\scriptsize{$\pm$0.16}} & 0.94{\scriptsize{$\pm$0.12}} & 1.23{\scriptsize{$\pm$0.15}} & 1.17{\scriptsize{$\pm$0.13}} & 0.32{\scriptsize{$\pm$0.08}} & $<$0.19{\scriptsize{$\pm$0.08}} \\
0.10--0.15 & 2.49{\scriptsize{$\pm$0.26}} & 2.37{\scriptsize{$\pm$0.25}} & 1.36{\scriptsize{$\pm$0.15}} & 1.23{\scriptsize{$\pm$0.13}} & 1.41{\scriptsize{$\pm$0.15}} & 1.33{\scriptsize{$\pm$0.14}} & 0.45{\scriptsize{$\pm$0.07}} & 0.43{\scriptsize{$\pm$0.07}} \\
0.15--0.20 & 2.13{\scriptsize{$\pm$0.23}} & 3.21{\scriptsize{$\pm$0.34}} & 2.26{\scriptsize{$\pm$0.24}} & 2.23{\scriptsize{$\pm$0.23}} & 1.88{\scriptsize{$\pm$0.20}} & 1.79{\scriptsize{$\pm$0.20}} & 1.00{\scriptsize{$\pm$0.13}} & 0.85{\scriptsize{$\pm$0.12}} \\
0.20--0.50 & 4.85{\scriptsize{$\pm$0.49}} & 4.91{\scriptsize{$\pm$0.50}} & 4.13{\scriptsize{$\pm$0.42}} & 2.95{\scriptsize{$\pm$0.30}} & 3.21{\scriptsize{$\pm$0.33}} & 3.12{\scriptsize{$\pm$0.32}} & 1.29{\scriptsize{$\pm$0.15}} & 1.48{\scriptsize{$\pm$0.17}} \\
0.50--1.00 & 11.42{\scriptsize{$\pm$1.15}} & 11.49{\scriptsize{$\pm$1.16}} & 8.30{\scriptsize{$\pm$0.85}} & 6.41{\scriptsize{$\pm$0.66}} & 6.06{\scriptsize{$\pm$0.65}} & 3.45{\scriptsize{$\pm$0.41}} & 4.66{\scriptsize{$\pm$0.51}} & 4.17{\scriptsize{$\pm$0.47}} \\
1.00--3.00 & 24.35{\scriptsize{$\pm$2.44}} & 19.39{\scriptsize{$\pm$1.97}} & 17.80{\scriptsize{$\pm$1.82}} & 9.40{\scriptsize{$\pm$1.00}} & 15.93{\scriptsize{$\pm$1.68}} & 5.68{\scriptsize{$\pm$0.68}} & 10.01{\scriptsize{$\pm$1.11}} & 6.28{\scriptsize{$\pm$0.72}} \\
\enddata
\end{deluxetable*}

\begin{deluxetable*}{ccccccccc}
\tabletypesize{\footnotesize}
\tablecaption{Average 160~$\mu$m Flux Densities [mJy] and Errors \label{tab:160}}
\tablehead{\colhead{$S_{24}$} & \multicolumn{8}{c}{Redshift Range} \\
\colhead{(mJy)} & \colhead{0--0.4} & \colhead{0.4--0.6} & \colhead{0.6--0.8} & \colhead{0.8--1.0} & \colhead{1.0--1.2} & \colhead{1.2--1.6} & \colhead{1.6--2} & \colhead{2--3} } 
\startdata
0.06--0.08 & 6.19{\scriptsize{$\pm$0.68}} & 2.99{\scriptsize{$\pm$0.43}} & 3.03{\scriptsize{$\pm$0.37}} & 2.42{\scriptsize{$\pm$0.33}} & 1.93{\scriptsize{$\pm$0.29}} & 4.94{\scriptsize{$\pm$0.53}} & $<$2.29{\scriptsize{$\pm$0.32}} & $<$1.87{\scriptsize{$\pm$0.30}} \\
0.08--0.10 & 5.09{\scriptsize{$\pm$0.61}} & 5.01{\scriptsize{$\pm$0.61}} & 4.02{\scriptsize{$\pm$0.50}} & 4.60{\scriptsize{$\pm$0.53}} & 4.23{\scriptsize{$\pm$0.50}} & 4.15{\scriptsize{$\pm$0.49}} & 1.88{\scriptsize{$\pm$0.36}} & $<$2.79{\scriptsize{$\pm$0.40}} \\
0.10--0.15 & 7.70{\scriptsize{$\pm$0.83}} & 7.25{\scriptsize{$\pm$0.78}} & 8.48{\scriptsize{$\pm$0.88}} & 7.00{\scriptsize{$\pm$0.73}} & 5.10{\scriptsize{$\pm$0.55}} & 6.96{\scriptsize{$\pm$0.73}} & 4.44{\scriptsize{$\pm$0.48}} & 4.90{\scriptsize{$\pm$0.56}} \\
0.15--0.20 & 8.11{\scriptsize{$\pm$0.88}} & 8.34{\scriptsize{$\pm$0.93}} & 8.92{\scriptsize{$\pm$0.95}} & 6.28{\scriptsize{$\pm$0.69}} & 9.92{\scriptsize{$\pm$1.05}} & 10.22{\scriptsize{$\pm$1.07}} & 6.20{\scriptsize{$\pm$0.69}} & 6.10{\scriptsize{$\pm$0.72}} \\
0.20--0.50 & 15.07{\scriptsize{$\pm$1.52}} & 15.62{\scriptsize{$\pm$1.60}} & 15.16{\scriptsize{$\pm$1.54}} & 12.74{\scriptsize{$\pm$1.30}} & 11.47{\scriptsize{$\pm$1.18}} & 17.50{\scriptsize{$\pm$1.79}} & 6.95{\scriptsize{$\pm$0.73}} & 9.92{\scriptsize{$\pm$1.05}} \\
0.50--1.00 & 28.72{\scriptsize{$\pm$2.91}} & 27.14{\scriptsize{$\pm$2.78}} & 30.95{\scriptsize{$\pm$3.17}} & 24.62{\scriptsize{$\pm$2.55}} & 19.85{\scriptsize{$\pm$2.19}} & 17.57{\scriptsize{$\pm$2.03}} & 18.88{\scriptsize{$\pm$2.09}} & 17.19{\scriptsize{$\pm$1.89}} \\
1.00--3.00 & 51.58{\scriptsize{$\pm$5.21}} & 51.28{\scriptsize{$\pm$5.30}} & 39.69{\scriptsize{$\pm$4.13}} & 26.11{\scriptsize{$\pm$3.01}} & 32.53{\scriptsize{$\pm$3.97}} & $<$16.16{\scriptsize{$\pm$2.07}} & 18.94{\scriptsize{$\pm$2.48}} & 21.76{\scriptsize{$\pm$2.71}} \\
\enddata
\end{deluxetable*}

\begin{deluxetable*}{ccccccccc}
\tabletypesize{\footnotesize}
\tablecaption{Average $L_{\rm{IR}} = L(8-1000 \mu$m) in [log $L_{\sun}$]  \label{tab:LIR}}
\tablehead{\colhead{} & \multicolumn{8}{c}{Redshift Range} \\
\colhead{$S_{24}$} & \colhead{0--0.4} & \colhead{0.4--0.6} & \colhead{0.6--0.8} & \colhead{0.8--1.0} & \colhead{1.0--1.2} & \colhead{1.2--1.6} & \colhead{1.6--2} & \colhead{2--3} \\
\colhead{(mJy)} & \colhead{(0.3)} & \colhead{(0.5)} & \colhead{(0.7)} & \colhead{(0.9)} & \colhead{(1.1)} & \colhead{(1.4)} & \colhead{(1.8)} & \colhead{(2.3)} }
\startdata
0.06--0.08 & 9.96{\scriptsize{$\pm$0.07}} & 10.41{\scriptsize{$\pm$0.09}} & 10.74{\scriptsize{$\pm$0.10}} & 10.90{\scriptsize{$\pm$0.09}} & 11.12{\scriptsize{$\pm$0.09}} & 11.66{\scriptsize{$\pm$0.09}} & $<$11.60{\scriptsize{$\pm$0.10}} & $<$11.84{\scriptsize{$\pm$0.12}} \\
0.08--0.10 & 9.97{\scriptsize{$\pm$0.07}} & 10.66{\scriptsize{$\pm$0.08}} & 10.89{\scriptsize{$\pm$0.07}} & 11.10{\scriptsize{$\pm$0.12}} & 11.39{\scriptsize{$\pm$0.09}} & 11.69{\scriptsize{$\pm$0.11}} & 11.62{\scriptsize{$\pm$0.12}} & $<$12.01{\scriptsize{$\pm$0.14}} \\
0.10--0.15 & 10.16{\scriptsize{$\pm$0.06}} & 10.82{\scriptsize{$\pm$0.07}} & 11.11{\scriptsize{$\pm$0.10}} & 11.27{\scriptsize{$\pm$0.12}} & 11.47{\scriptsize{$\pm$0.10}} & 11.84{\scriptsize{$\pm$0.09}} & 11.88{\scriptsize{$\pm$0.13}} & 12.22{\scriptsize{$\pm$0.18}} \\
0.15--0.20 & 10.18{\scriptsize{$\pm$0.07}} & 10.89{\scriptsize{$\pm$0.08}} & 11.19{\scriptsize{$\pm$0.08}} & 11.35{\scriptsize{$\pm$0.08}} & 11.69{\scriptsize{$\pm$0.10}} & 12.00{\scriptsize{$\pm$0.08}} & 12.06{\scriptsize{$\pm$0.11}} & 12.36{\scriptsize{$\pm$0.20}} \\
0.20--0.50 & 10.49{\scriptsize{$\pm$0.07}} & 11.16{\scriptsize{$\pm$0.08}} & 11.46{\scriptsize{$\pm$0.10}} & 11.61{\scriptsize{$\pm$0.10}} & 11.82{\scriptsize{$\pm$0.11}} & 12.24{\scriptsize{$\pm$0.17}} & 12.12{\scriptsize{$\pm$0.20}} & 12.55{\scriptsize{$\pm$0.22}} \\
0.50--1.00 & 10.81{\scriptsize{$\pm$0.07}} & 11.44{\scriptsize{$\pm$0.06}} & 11.77{\scriptsize{$\pm$0.10}} & 11.92{\scriptsize{$\pm$0.10}} & 12.07{\scriptsize{$\pm$0.11}} & 12.33{\scriptsize{$\pm$0.19}} & 12.55{\scriptsize{$\pm$0.15}} & 12.83{\scriptsize{$\pm$0.20}} \\
1.00--3.00 & 11.10{\scriptsize{$\pm$0.07}} & 11.69{\scriptsize{$\pm$0.07}} & 11.95{\scriptsize{$\pm$0.07}} & 11.98{\scriptsize{$\pm$0.09}} & 12.34{\scriptsize{$\pm$0.08}} & 12.23{\scriptsize{$\pm$0.11}} & 12.68{\scriptsize{$\pm$0.20}} & 13.04{\scriptsize{$\pm$0.23}} \\
\enddata
\tablecomments{Average redshifts and $S_{24}$ for each bin are given in parentheses.  Bins in which both 70 $\mu$m and 160 $\mu$m stacks resulted in non-detections are marked as upper limits.}
\end{deluxetable*}

\begin{figure*}[!h]
\figurenum{5b}
\centering
\includegraphics[width=6in]{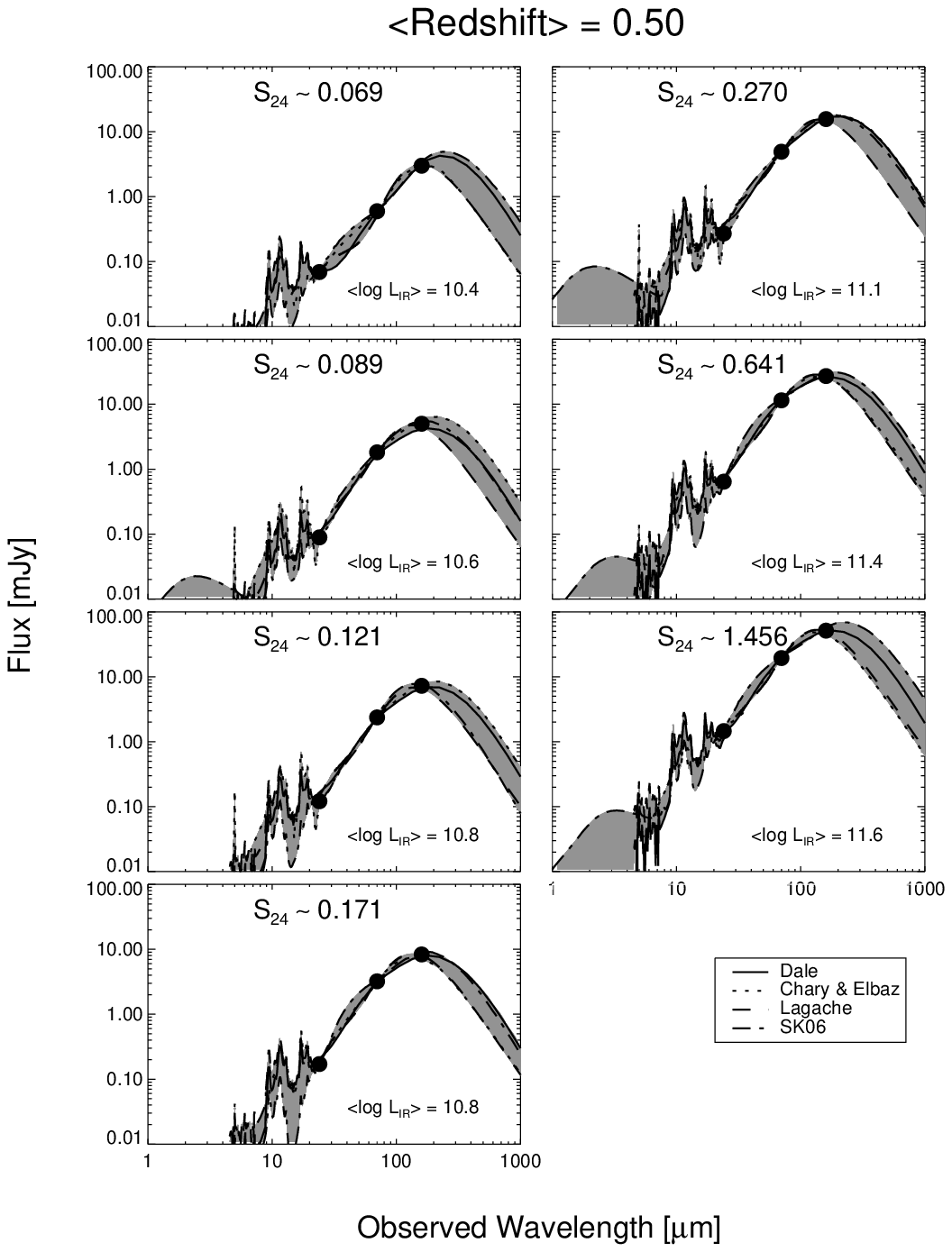}
\caption{SED fits to stacked fluxes in each $S_{24}$ bin at a redshift $0.4 \leq z \leq 0.6$.  The region spanned by the best-fit SEDs from each library is shaded.  Non-detections are marked as upper limits, and the errors on the rest of the points are smaller than the size of the dot. Libraries used are from \citeauthor{Dale:2002p1836} (solid line), \citeauthor{Chary:2001p1425} (dotted line), \citeauthor{Lagache:2003p1867} (dashed line), and \citeauthor{Siebenmorgen:2007p1876} (dot-dashed line).} 
\label{fig:matchz1}
\end{figure*}

\begin{figure*}[!h]
\figurenum{5c}
\centering
\includegraphics[width=6in]{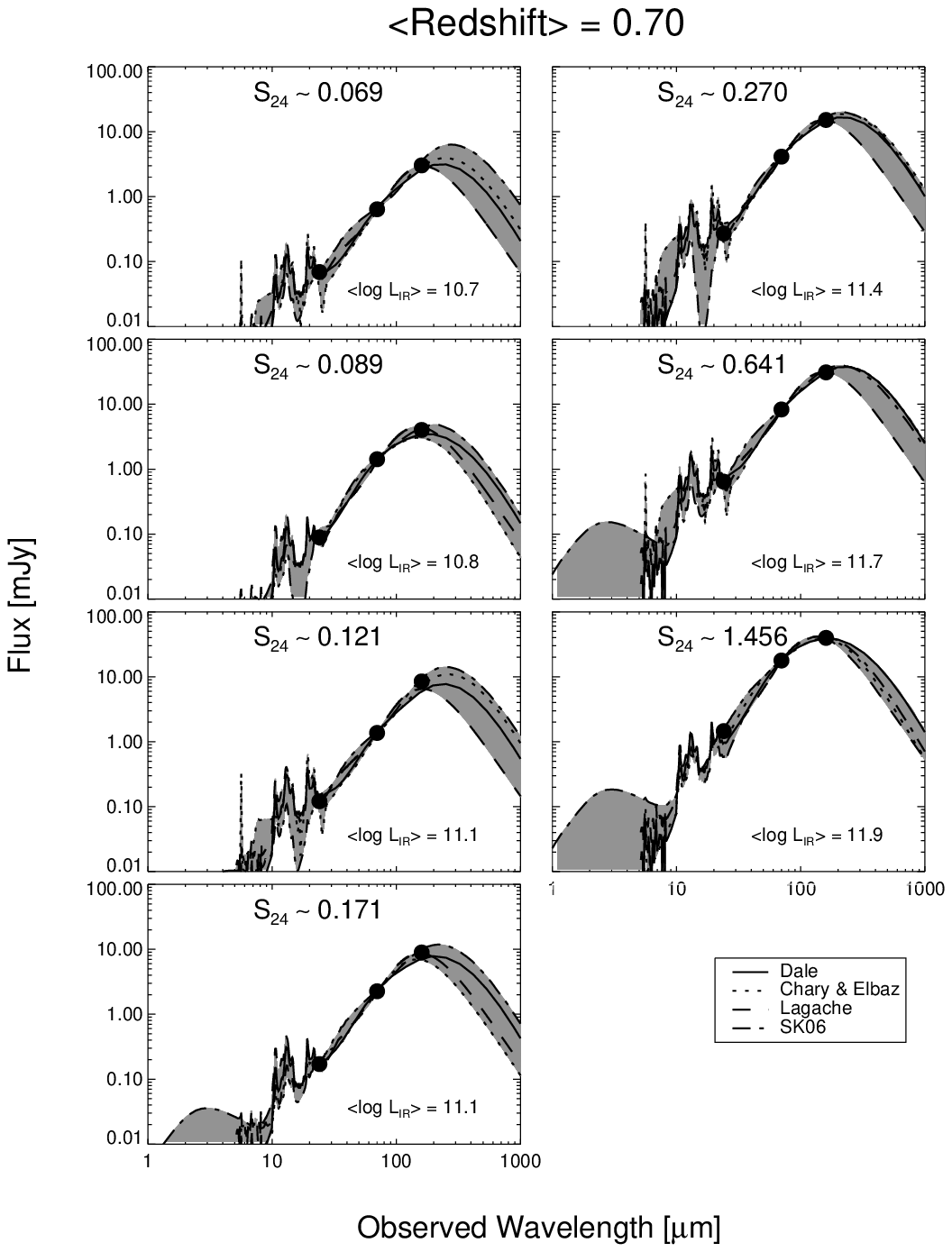}
\caption{SED fits to stacked fluxes in each $S_{24}$ bin at a redshift $0.6 \leq z \leq 0.8$.  The region spanned by the best-fit SEDs from each library is shaded.  Non-detections are marked as upper limits, and the errors on the rest of the points are smaller than the size of the dot. Libraries used are from \citeauthor{Dale:2002p1836} (solid line), \citeauthor{Chary:2001p1425} (dotted line), \citeauthor{Lagache:2003p1867} (dashed line), and \citeauthor{Siebenmorgen:2007p1876} (dot-dashed line).}
\label{fig:matchz2}
\end{figure*}

\begin{figure*}[!h]
\figurenum{5d}
\centering
\includegraphics[width=6in]{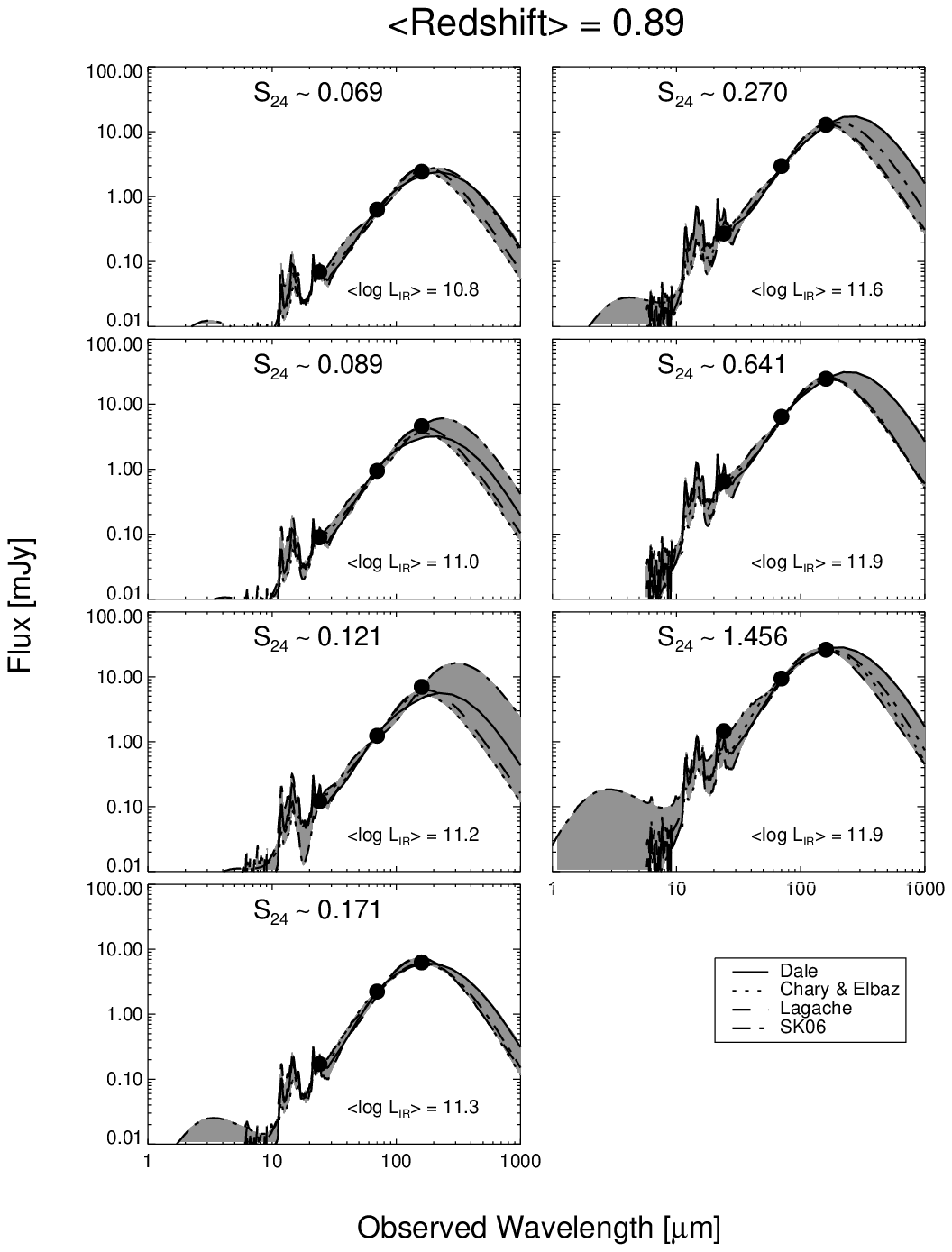}
\caption{SED fits to stacked fluxes in each $S_{24}$ bin at a redshift $0.8 \leq z \leq 1.0$.  The region spanned by the best-fit SEDs from each library is shaded.  Non-detections are marked as upper limits, and the errors on the rest of the points are smaller than the size of the dot. Libraries used are from \citeauthor{Dale:2002p1836} (solid line), \citeauthor{Chary:2001p1425} (dotted line), \citeauthor{Lagache:2003p1867} (dashed line), and \citeauthor{Siebenmorgen:2007p1876} (dot-dashed line).}
\label{fig:matchz3}
\end{figure*}

\begin{figure*}[!h]
\figurenum{5e}
\centering
\includegraphics[width=6in]{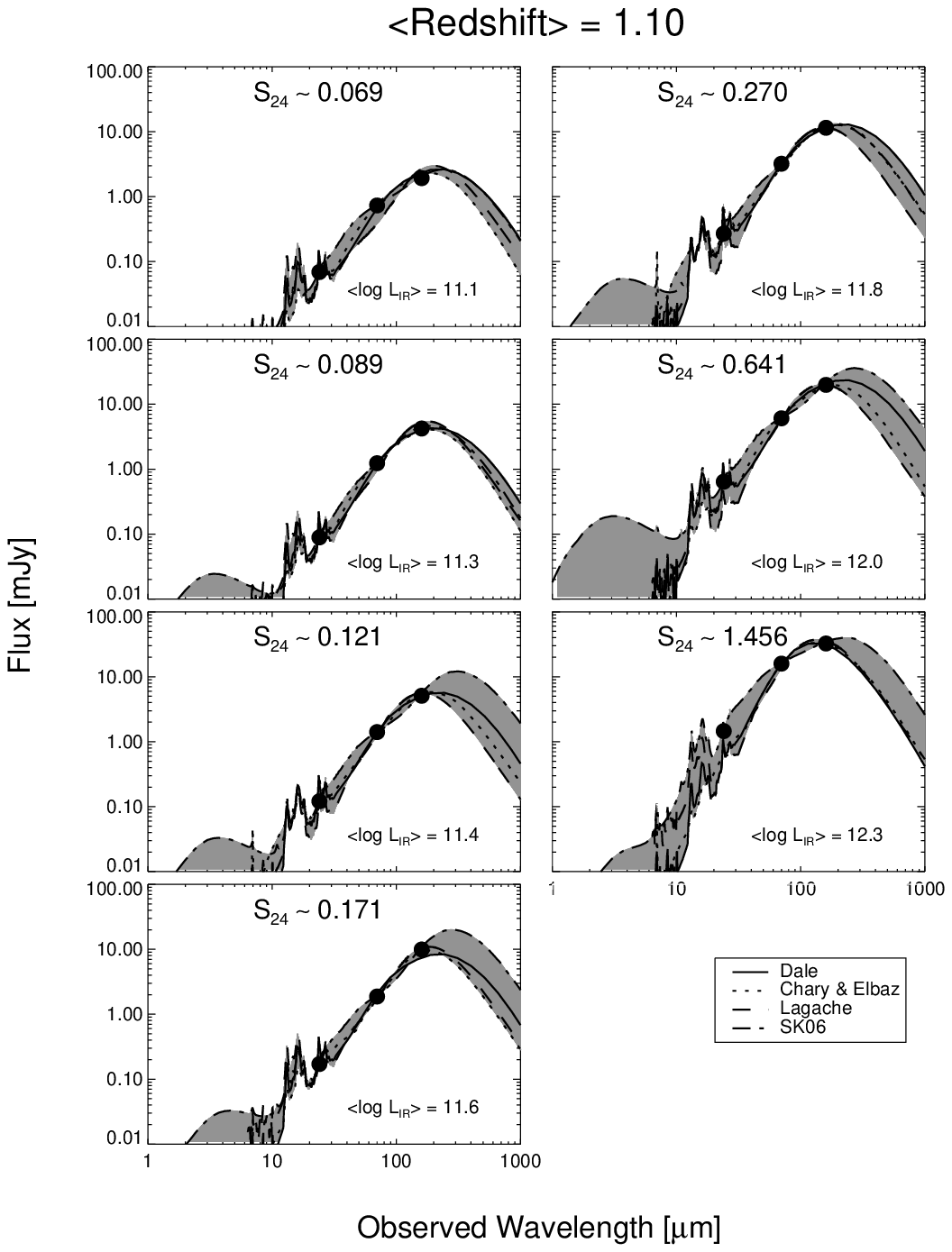}
\caption{SED fits to stacked fluxes in each $S_{24}$ bin at a redshift $1.0 \leq z \leq 1.2$.  The region spanned by the best-fit SEDs from each library is shaded.  Non-detections are marked as upper limits, and the errors on the rest of the points are smaller than the size of the dot. Libraries used are from \citeauthor{Dale:2002p1836} (solid line), \citeauthor{Chary:2001p1425} (dotted line), \citeauthor{Lagache:2003p1867} (dashed line), and \citeauthor{Siebenmorgen:2007p1876} (dot-dashed line).}
\label{fig:matchz4}
\end{figure*}

\begin{figure*}[!h]
\figurenum{5f}
\centering
\includegraphics[width=6in]{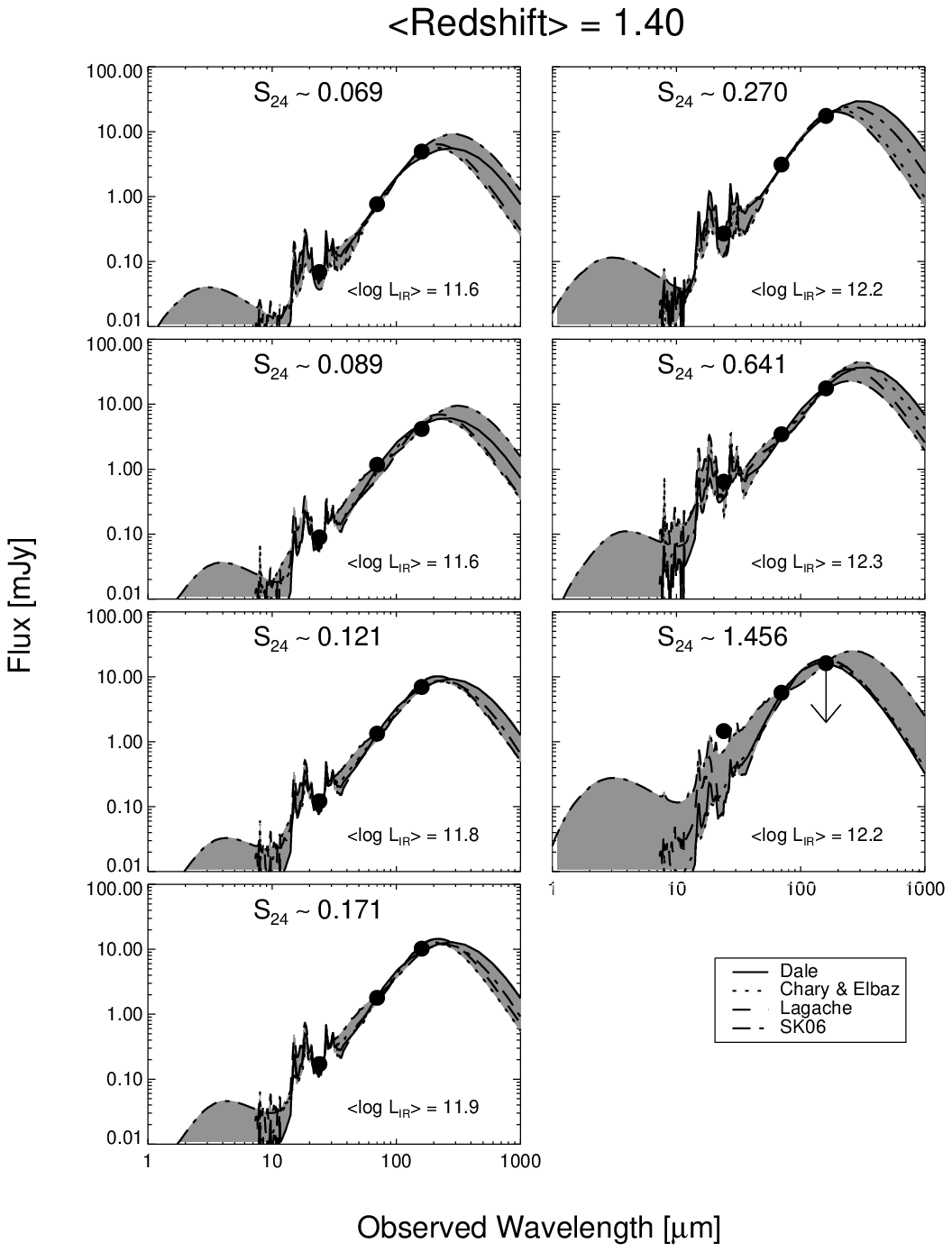}
\caption{SED fits to stacked fluxes in each $S_{24}$ bin at a redshift $1.2 \leq z \leq 1.6$.  The region spanned by the best-fit SEDs from each library is shaded.  Non-detections are marked as upper limits, and the errors on the rest of the points are smaller than the size of the dot. Libraries used are from \citeauthor{Dale:2002p1836} (solid line), \citeauthor{Chary:2001p1425} (dotted line), \citeauthor{Lagache:2003p1867} (dashed line), and \citeauthor{Siebenmorgen:2007p1876} (dot-dashed line).}
\label{fig:matchz5}
\end{figure*}

\begin{figure*}[!h]
\centering
\figurenum{5g}
\includegraphics[width=6in]{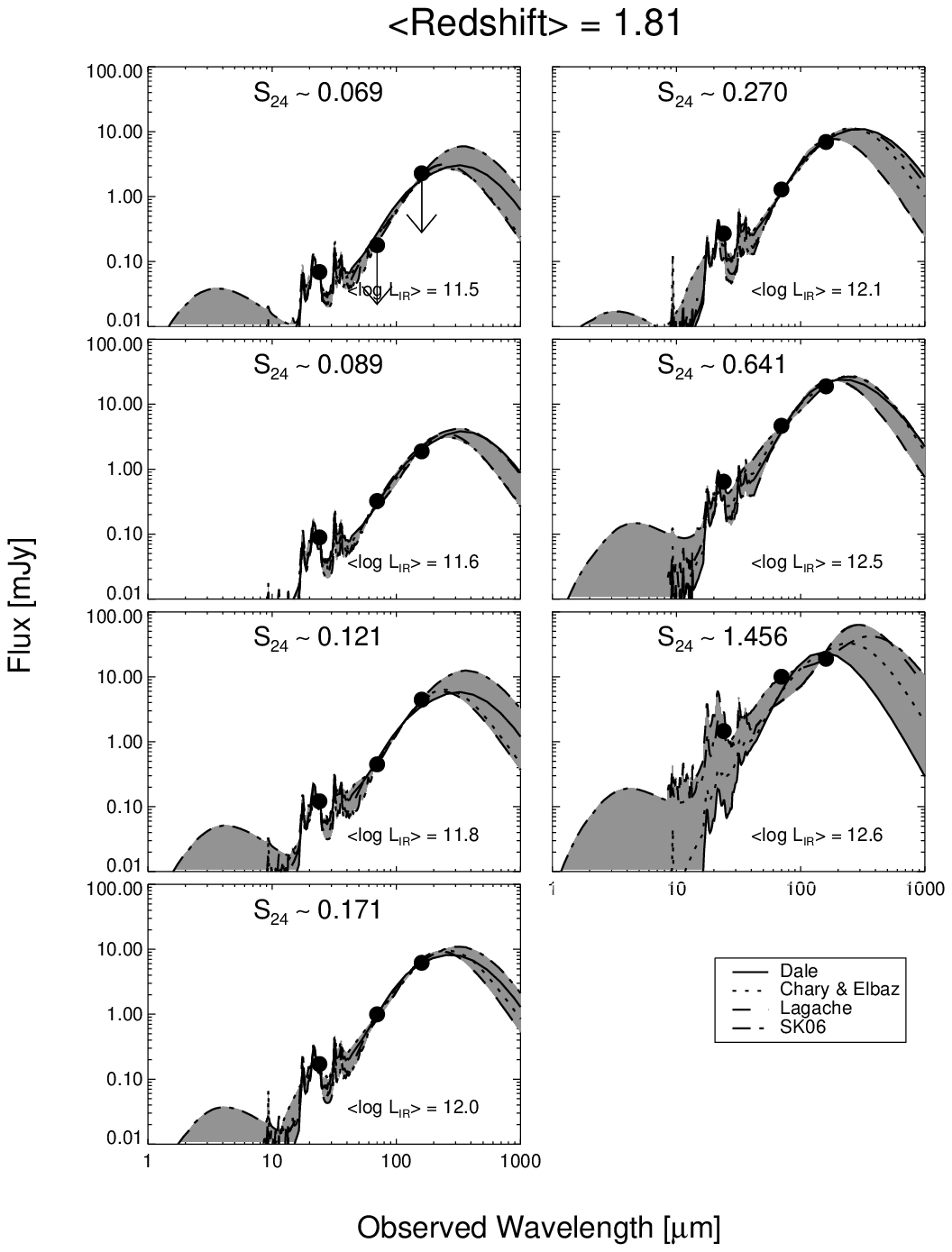}
\caption{SED fits to stacked fluxes in each $S_{24}$ bin at a redshift $1.6 \leq z \leq 2.0$.  The region spanned by the best-fit SEDs from each library is shaded.  Non-detections are marked as upper limits, and the errors on the rest of the points are smaller than the size of the dot. Libraries used are from \citeauthor{Dale:2002p1836} (solid line), \citeauthor{Chary:2001p1425} (dotted line), \citeauthor{Lagache:2003p1867} (dashed line), and \citeauthor{Siebenmorgen:2007p1876} (dot-dashed line).}
\label{fig:matchz6}
\end{figure*}

\begin{figure*}[!h]
\centering
\figurenum{5h}
\includegraphics[width=6in]{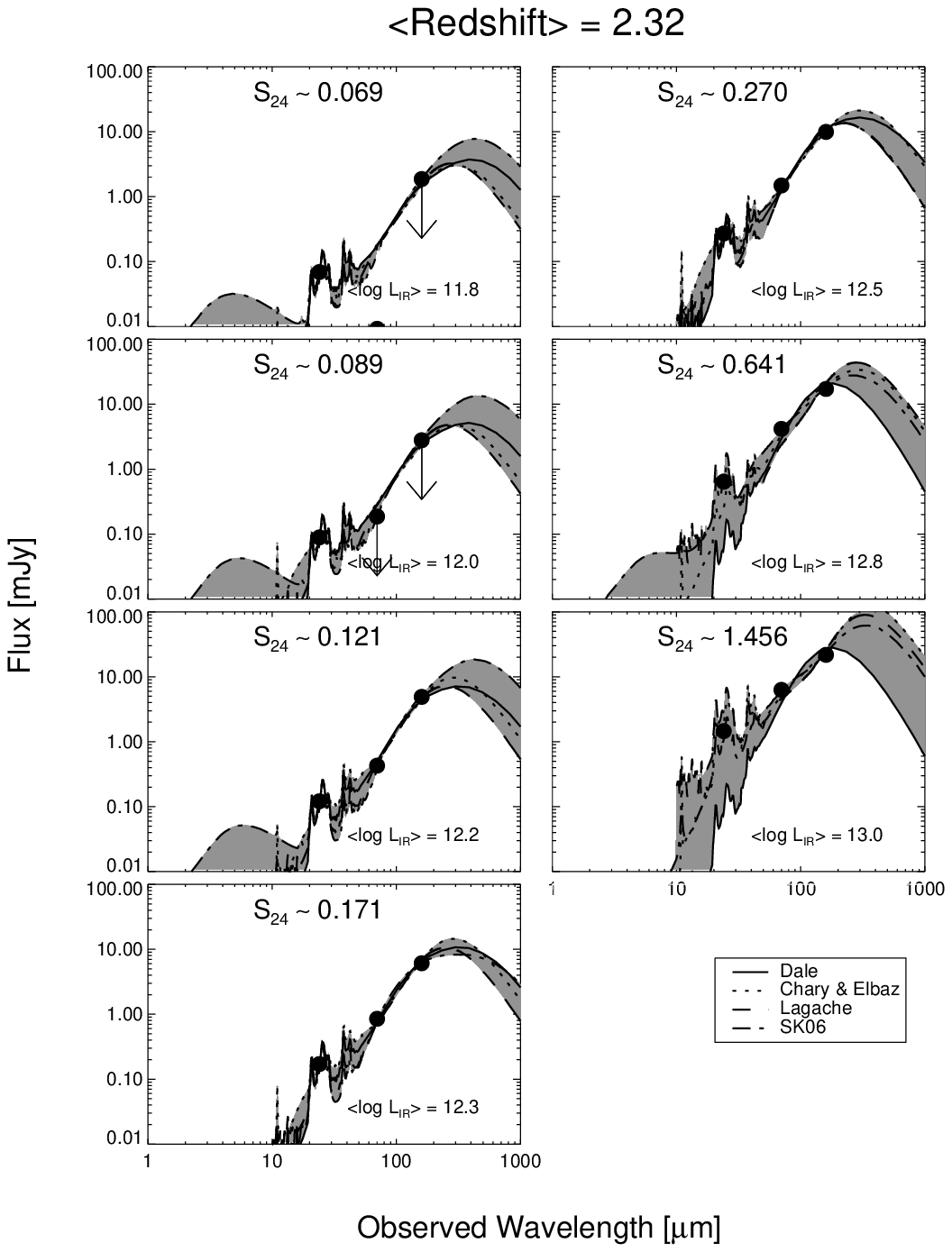}
\caption{SED fits to stacked fluxes in each $S_{24}$ bin at a redshift $2.0 \leq z \leq 3.0$.  The region spanned by the best-fit SEDs from each library is shaded.  Non-detections are marked as upper limits, and the errors on the rest of the points are smaller than the size of the dot. Libraries used are from \citeauthor{Dale:2002p1836} (solid line), \citeauthor{Chary:2001p1425} (dotted line), \citeauthor{Lagache:2003p1867} (dashed line), and \citeauthor{Siebenmorgen:2007p1876} (dot-dashed line).}
\label{fig:matchz7}
\end{figure*}

\begin{figure*}[!h]
\centering
\figurenum{6}
\plotone{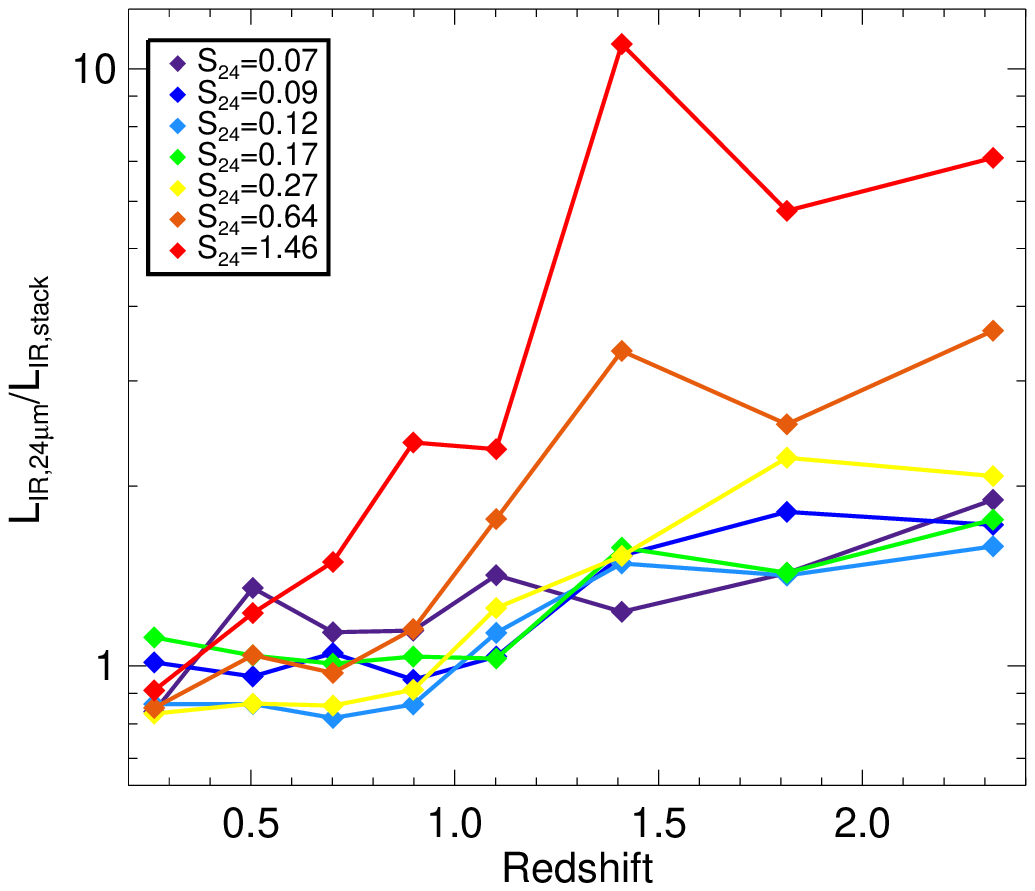}
\caption{Ratio of $L_{\rm{IR}}$ derived from only using $S_{24}$ vs. $L_{\rm{IR}}$ derived from our stacking analysis.  There is fairly good agreement between the two methods at low redshifts, but at higher redshifts we see that previous methods using only 24 $\mu$m flux overestimate the true luminosity, especially in the brighter 24 $\mu$m flux bins.}
\label{fig:matchLIR}
\end{figure*}


\clearpage
\bibliographystyle{apj}
\bibliography{stacking_final,apj-jour}

\end{document}